\documentclass[10pt,aps,prd,floatfix,notitlepage,nofootinbib,superscriptaddress]{revtex4-1}

\usepackage{latexsym}
\usepackage{epsfig}
\usepackage{epstopdf}
\usepackage{graphicx}
\usepackage{amssymb}
\usepackage{amsmath}
\usepackage{dcolumn}
\usepackage{bm}
\usepackage{color}

\begin{document}

\title{Interacting quintessence from a variational approach \\ Part II: derivative couplings}

\author{Christian G. B\"ohmer}
\email{c.boehmer@ucl.ac.uk}
\affiliation{Department of Mathematics, University College London, Gower Street, London, WC1E 6BT, UK}

\author{Nicola Tamanini}
\email{nicola.tamanini@cea.fr}
\affiliation{Institut de Physique Th{\'e}orique, CEA-Saclay, F-91191, Gif-sur-Yvette, France}

\author{Matthew Wright}
\email{matthew.wright.13@ucl.ac.uk}
\affiliation{Department of Mathematics, University College London, Gower Street, London, WC1E 6BT, UK}

\begin{abstract}
We consider an original variational approach for building new models of quintessence interacting with dark or baryonic matter. The coupling is introduced at the Lagrangian level using a variational formulation for relativistic fluids, where the interacting term generally depends on both the dynamical degrees of freedom of the theory and their spacetime derivatives. After deriving the field equations from the action, we consider applications in the context of cosmology. Two simple models are studied using dynamical system techniques showing the interesting phenomenology arising in this framework. We find that these models contain dark energy dominated late time attractors with early time matter dominated epochs and also obtain a possible dynamical crossing of the phantom barrier. The formulation and results presented here complete and expand the analysis exposed in the first part of this work, where only algebraic couplings, without spacetime derivatives, were considered.
\end{abstract}

\maketitle

\tableofcontents

\section{Introduction}

The aim of the present work is to expand and complete the analysis performed in \cite{part1}, which hereafter will be referred to as Part~I of this study. In Part~I new models of interacting dark energy have been introduced starting from a variational approach. Defining a consistent variational set up for interacting dark energy models is an important issue not only for exploring new physical models in late time cosmology, but also for producing new theoretical mechanisms responsible for the phenomenology of such interactions. From the observational perspective a coupling between dark energy and dark matter might give rise to some effects which could be detected in forthcoming surveys. In fact, although non-interacting dark energy models can still accommodate the observations \cite{Komatsu:2010fb}, such interaction is mildly favoured by the data \cite{Pettorino:2013oxa}, and can even alleviate some tensions present in the comparison of different datasets \cite{Salvatelli:2013wra}. For these reasons the theoretical development of interacting dark energy models should proceed to both drive future observations toward possible detectable features and to justify all the phenomenological models considered so far in the literature; see e.g.~\cite{Bolotin:2013jpa} and references therein.

The particular advantage arising from a variational formulation consists in generating fully covariant equations of motion, which can then be applied to cosmology at both the background and the perturbation level. It is a well known problem to understand how to promote the phenomenological background equations of interacting dark energy to their covariant counterparts, and a satisfactory solution is still missing \cite{Koivisto:2005nr}. In fact, it will always be possible to construct two covariant theories which are equivalent at the background level but will have different perturbations. This is due to the non-linearity of the Einstein field equations in general. Hence, there is a strong motivation to construct models at the level of the action.

A variational approach, such as the one advanced in Part I, permits to solve this problem by automatically producing the desired field equations. In Part I a canonical scalar field describing dark energy, namely quintessence, has been coupled to a barotropic fluid at the Lagrangian level; see also \cite{Pourtsidou:2013nha} for similar ideas. The Lagrangian description of a relativistic fluid that has been used in Part I, and which will also be employed here, is the one outlined by Brown in \cite{Brown:1992kc}. In what follows we will not review the features of this formulation but refer the reader to Part~I. All details and the thermodynamics are thoroughly discussed in Brown's original paper.

The main issue developed and studied in this Part~II, is the possibility of directly coupling the 4-velocity $U^\mu$ of a relativistic matter fluid to a scalar field $\phi$. The simplest way of mixing $U^\mu$ and $\phi$ in an interacting Lagrangian is through the scalar $U^\mu\partial_\mu\phi$ which inevitably involves the use of a spacetime derivative. In Part~I we considered only {\it algebraic couplings} between the scalar field and the fluid's degrees of freedom where no spacetime derivatives are allowed. This in particular implies that the coupling $U^\mu\partial_\mu\phi$ falls beyond the interacting quintessence models introduced and investigated there. The scope of this Part~II is exactly to complete the analysis of Part~I by studying {\it derivative couplings}, i.e.~coupling where spacetime derivatives are allowed.

The paper is organized as follows. In Sec.~\ref{sec:derivative_coupling} we will introduce the Lagrangian of a scalar field interacting with a relativistic perfect fluid through a derivative coupling. Brown's formulation will be employed to define and treat the coupling terms which will be assumed to depend on as few spacetime derivatives as possible in alignment with the effective field theory approach. In Sec.~\ref{sec:cosmology} we will then focus on cosmological applications, finding first the relevant equations governing the evolution of the universe at large scales, and then analysing them with dynamical systems techniques. Finally in Sec.~\ref{sec:conclusion} we will discuss the results obtained, highlighting the differences between the algebraic couplings of Part~I and the derivative couplings of Part~II. In Sec.~\ref{sec:conclusion} we will also draw conclusions and speculate on possible future works.

{\bf Notation and conventions}: Unless otherwise specified we will assume standard general relativistic notation with the metric convention $(-1,1,1,1)$ and Greek indices running from 0 to 3. Sometimes the comma notation for partial derivatives will be used: for example $\phi_{,\mu}=\partial_\mu\phi$. Units where $c=\hbar=1$ will be employed together with the definitions $\kappa^2=M_{\rm P}^{-2}=8\pi G$.

\section{Relativistic fluid interacting with a scalar field: derivative coupling}
\label{sec:derivative_coupling}

In this section we generalize the algebraic couplings considered in the first part of this work \cite{part1} by allowing also terms depending on the scalar field's first derivative $\partial_\mu\phi$ to interact with the fluid's degrees of freedom. In analogy to the effective field theory approach, we will only consider couplings with as few spacetime derivatives as possible. Moreover, we will not consider non-canonical couplings to the scalar field's kinetic term as this is beyond the scope of the present work. However, it would be a possible further extension of the variational approach to also discuss such models.

\subsection{Lagrangian formulation and field equations}

The total action of our interacting dark energy system is \cite{part1}
\begin{align}
\mathcal{S} = \int d^4x \left(\mathcal{L}_{\rm grav} +\mathcal{L}_M+ \mathcal{L}_\phi+ \mathcal{L}_{\rm int}\right) \,,
\label{010b}
\end{align}
where the gravitational sector $\mathcal{L}_{\rm grav}$ is given by the standard Einstein-Hilbert Lagrangian
\begin{align}
\mathcal{L}_{\rm grav} = \frac{\sqrt{-g}}{2\kappa^2}R \,,
\end{align}
with $R$ being the curvature scalar with respect to the metric $g_{\mu\nu}$ and $g$ its determinant. The Lagrangian of the scalar field is assumed to be in the canonical form
\begin{align}
\mathcal{L}_\phi = -\sqrt{-g}\, \left[\frac{1}{2}\partial_\mu\phi\,\partial^\mu\phi +V(\phi)\right] \,,
\end{align}
where $V$ is an arbitrary potential depending on $\phi$. Using Brown's formulation \cite{part1,Brown:1992kc} the Lagrangian for the relativistic fluid $\mathcal{L}_M$ can be written as
\begin{align}
\mathcal{L}_M = -\sqrt{-g}\,\rho(n,s) + J^\mu\left(\varphi_{,\mu}+s\theta_{,\mu}+\beta_A\alpha^A_{,\mu}\right) \,,
\label{001b}
\end{align}
where $\rho$ is the energy density of the fluid prescribed as a function of $n$, the particle number density, and $s$, the entropy density per particle. The fields $\varphi$, $\theta$ and $\beta_A$ are all Lagrange multipliers with $A$ taking the values $1,2,3$ and $\alpha_A$ are the Lagrangian coordinates of the fluid. The vector-density particle number flux $J^\mu$ is related to $n$ as
\begin{align}
J^\mu=\sqrt{-g}\,n\,U^\mu\,, \qquad |J|=\sqrt{-g_{\mu\nu}J^\mu J^\nu}\,, \qquad n=\frac{|J|}{\sqrt{-g}} \,,
\label{056}
\end{align}
where $U^\mu$ is the fluid 4-velocity satisfiyng $U_\mu U^\mu=-1$.
For further details regarding Brown's Lagrangian formalism for relativistic fluids we refer to \cite{Brown:1992kc}. 

Considering the dynamical degrees of freedom of the above Lagrangian, we note that the only two possible scalar coupling terms (up to total derivatives) that can be constructed containing only first order derivatives of the scalar field $\phi$ are $\partial\phi^2=g^{\mu\nu}\partial_\mu\phi\partial_\nu\phi$ and $J^\mu\partial_\mu\phi$. The first possibility is nothing but the usual kinetic term for the scalar field, while the second one is a new coupling term that can only be considered within the fluid Lagrangian formalism treated in this work. In particular we will only study a linear coupling to $J^\mu\partial_\mu\phi$ leaving higher order couplings aside at present.
Note that within an effective field theory framework such higher order terms would naturally be neglected at first order.
Therefore, the coupling term we will study in what follows is
\begin{align}
\mathcal{L}_{\rm int} = f(n,s,\phi) J^\mu\partial_\mu\phi \,,
\label{302b}
\end{align}
where $f$ is an arbitrary function.
This is the most general coupling term where only one spacetime derivative appears.

The variations of (\ref{010b}) with respect to the Lagrange multipliers $\varphi$, $\theta$, $\beta_A$ and Lagrangian coordinates of the fluid $\alpha^A$ give the equations
\begin{align}
\varphi:& \qquad J^\mu{}_{,\mu}=0 \,,\label{004b}\\
\theta:& \qquad (sJ^\mu)_{,\mu}=0 \,,\label{005b}\\
\beta_A:& \qquad J^\mu\alpha^A_{,\mu}=0 \,,\label{006b}\\
\alpha^A:& \qquad (J^\mu\beta_A)_{,\mu}=0 \label{007b}\,,
\end{align}
which are not modified by the coupling with the scalar field; see \cite{part1,Brown:1992kc}.
Eqs.~(\ref{004b}) and (\ref{005b}) stand for the particle number conservation constraint and the entropy exchange constraint, respectively. These can be rewritten as
\begin{align}
\nabla_\mu(n\,U^\mu)=0 \quad{\rm and}\quad \nabla_\mu(s\,n\,U^\mu)=0 \,,
\label{019b}
\end{align}
where $\nabla_\mu$ is the covariant derivative with respect to $g_{\mu\nu}$. Eqs.~(\ref{006b}) and (\ref{007b}) determine the dynamics of the Lagrange multipliers but are not needed for our scopes and will not be considered further in what follows; see \cite{Brown:1992kc} for more information about their physical meaning. The variations with respect to $J^\mu$ and $s$ yield, respectively
\begin{align}
\mu U_\mu +\varphi_{,\mu}+ s\theta_{,\mu} +\beta_A\alpha^A_{,\mu} &= \left(n\frac{\partial f}{\partial n} U_\mu U^\nu-f\delta^\nu_\mu\right)\phi_{,\nu} \,, \label{046b} \\
T&=U^\mu \left(\theta_{,\mu}+\frac{\partial f}{\partial s}\phi_{,\mu}\right) \,, \label{301b}
\end{align}
where the chemical potential $\mu$ and temperature $T$ are defined by
\begin{align}
\mu = \frac{\partial\rho}{\partial n} \quad\mbox{and}\quad T = \frac{1}{n}\frac{\partial\rho}{\partial n} \,, 
\label{011b}
\end{align}
Eqs.~(\ref{046b}) and (\ref{301b}) show how the four-velocity fluid decomposition and the temperature of the fluid depend also on the scalar field $\phi$ due to the non-vanishing interaction.
Similar equations are found also for the algebraic couplings considered in Part~I.
They modify the thermodynamics properties of the fluid reflecting the fact that an interaction with the scalar fields is now at play.

Variation of (\ref{010b}) with respect to $g_{\mu\nu}$ give the following Einstein field equations
\begin{align}
\frac{1}{\kappa^2}G_{\mu\nu} = T_{\mu\nu} +T_{\mu\nu}^{(\phi)} +T_{\mu\nu}^{\rm (int)} \,,
\label{043b}
\end{align}
where
\begin{align}
T_{\mu\nu} & = p\, g_{\mu\nu} + (\rho+p)\, U_\mu U_\nu \,, \\
T_{\mu\nu}^{(\phi)} &= \partial_\mu\phi\,\partial_\nu\phi -g_{\mu\nu} \left[\frac{1}{2}\partial_\mu\phi\,\partial^\mu\phi +V(\phi)\right] \,, \label{303b} \\
T_{\mu\nu}^{\rm (int)} &= -n^2\frac{\partial f}{\partial n} U^\lambda\partial_\lambda\phi \left(g_{\mu\nu}+U_\mu U_\nu\right) \,.
\label{030b}
\end{align}
with the fluid pressure defined as
\begin{align}
p = n\frac{\partial\rho}{\partial n}-\rho \,.
\label{eq:111}
\end{align}
Interestingly, now the energy-momentum tensor of the interaction is orthogonal to the fluid flow, $U^\mu T_{\mu\nu}^{\rm (int)}=0$ and it can be rewritten in a perfect fluid form
\begin{align}
	T_{\mu\nu}^{\rm (int)} = p_{\rm int}\,g_{\mu\nu} + \left(p_{\rm int}+\rho_{\rm int}\right) U_\mu U_\nu \,,
\end{align}
with vanishing energy density and pressure given by
\begin{align}
	\rho_{\rm int}=0 \qquad\mbox{and}\qquad p_{\rm int} =  -n^2\frac{\partial f}{\partial n} U^\lambda\partial_\lambda\phi \,.
	\label{996}
\end{align}
The scalar field equation, obtained from the variation of (\ref{010b}) with respect to $\phi$, is
\begin{align}
\Box\phi-V'+n^2\frac{\partial f}{\partial n} \nabla_\mu U^\mu = 0 \,.
\label{044b}
\end{align}
Note that whenever $f$ does not depend on the particle number density $n$ the interaction term does not give any contribution to the equations of motion since $T_{\mu\nu}^{\rm (int)}$ vanishes and Eq.~(\ref{044b}) reduces to the uncoupled Klein-Gordon equation. This is expected since in this case conditions (\ref{004b}) and (\ref{005b}) would constrain the interacting term (\ref{302b}) to become a total derivative and thus a boundary term. Note also that the interaction contributes always in the form $n^2\partial f/\partial n$ which does not depend on $n$ if one chooses $f \propto 1/n$. This case is particularly interesting since such a coupling would correspond to $\mathcal{L}_{\rm int} = \sqrt{-g}W(\phi)U^\mu\partial_\mu\phi$ with $W(\phi)$ being a general function of $\phi$ (ignoring the dependence on $s$). Its cosmological consequences will be studied in detail in Sec.~\ref{sec:cosmology}.

Intuitively, this result is not surprising. Our coupling is based on the idea of allowing an exchange of energy between the scalar field and the matter via the particle number density $n$ of that matter. Removing this dependence from the interaction Lagrangian corresponds to closing this channel of energy transfer, which, given that the fluid 4-velocity is conserved, is precisely why (\ref{302b}) becomes a boundary term in that case.

The original gravitational field equations that have been derived in this section, namely Eqs.~(\ref{043b}) and (\ref{044b}), represent a completely new way to couple a relativistic perfect fluid to a scalar field. Even when compared to the interacting equations obtained in \cite{part1}, one can realize that these equations not only take into account the coupling between $\phi$, $n$ and $s$, but heavily depend also on the fluid 4-velocity. These equations can be used to build new models of interacting quintessence (and coupled inflation) possibly leading to unexplored phenomenological and theoretical issues.

\subsection{Conservation equations}

In the following we will briefly discuss how our formalism compares with the standard approach of introducing couplings at the level of the field equations (see \cite{Tamanini:2015iia} for a general discussion on this issue). One can rewrite the coupled field equations above in the more familiar form
\begin{align}
  G_{\mu\nu} = \kappa^2 \left( T^{(A)}_{\mu\nu} + T^{(B)}_{\mu\nu} \right) \,, \qquad
  \nabla^\mu T^{(A)}_{\mu\nu} = Q_\nu \qquad\mbox{and}\qquad \nabla^\mu T^{(B)}_{\mu\nu} = -Q_\nu \,,
  \label{032b}
\end{align}
which is commonly used to couple two matter components in general relativity. For this purpose we define a new energy-momentum tensor for the fluid as $\tilde{T}_{\mu\nu} = T_{\mu\nu} + T_{\mu\nu}^{\rm (int)}$ with the energy density $\tilde\rho = \rho + \rho_{\rm int}$ and pressure $\tilde{p} = p + p_{\rm int}$.
The Einstein field equations (\ref{043b}) then become
\begin{align}
  G_{\mu\nu} = \kappa^2 \left(\tilde{T}_{\mu\nu}+T_{\mu\nu}^{(\phi)}\right)\,,
\end{align}
resembling the first of Eqs.~(\ref{032b}).
Using the Klein-Gordon equation (\ref{044b}) and the energy-momentum tensor (\ref{303b}), the conservation equation for the scalar field can be written as
\begin{align}
  \nabla^\mu T_{\mu\nu}^{(\phi)} = -n^2 \frac{\partial f}{\partial n} \nabla_\lambda U^\lambda \nabla_\nu\phi = Q_\nu\,,
  \label{048b}
\end{align}
which shows that the scalar field is not conserved due to the interaction with the fluid.

To prove that also the fluid energy-momentum is not conserved, it is better to split its conservation equation into the parallel and perpendicular components to the fluid flow:
\begin{align}
  \nabla^\mu \tilde{T}_{\mu\nu} = h^\lambda_\nu \nabla^\mu \tilde{T}_{\mu\lambda} - U_\nu U^\lambda \nabla^\mu \tilde{T}_{\mu\lambda} \,, \label{047b}
\end{align}
where $h_{\mu\nu}$ is defined by
\begin{align}
  h_{\mu\nu} = g_{\mu\nu} +U_\mu U_\nu \,.
  \label{017b}
\end{align}
Then using Eqs.~(\ref{004b})-(\ref{007b}) and Eq.~(\ref{046b}) one can show that (see the next subsection)
\begin{align}
  h_\mu^\nu \nabla^\lambda \tilde{T}_{\lambda\nu} = h_\mu^\nu n^2 \frac{\partial f}{\partial n} \nabla_\lambda U^\lambda \nabla_\nu\phi
  \qquad\mbox{and}\qquad
  U_\nu \nabla_\mu \tilde{T}^{\mu\nu} = U^\nu n^2 \frac{\partial f}{\partial n} \nabla_\lambda U^\lambda \nabla_\nu\phi \,,
  \label{053b}
\end{align}
which inserted back into Eq.~(\ref{047b}) gives
\begin{align}
  \nabla^\mu \tilde{T}_{\mu\nu} = n^2 \frac{\partial f}{\partial n} \nabla_\lambda U^\lambda \partial_\nu\phi = -Q_\nu\,.
  \label{049b}
\end{align}
This shows that the interacting approach developed in the present paper can be mapped back into the standard relativistic coupling between two matter components as given by Eqs.~(\ref{032b}) with an exchange vector $Q_\mu$ defined in Eqs.~(\ref{048b}) and (\ref{049b}). Note however that the pressure of the coupled fluid $\tilde{p}$ now differs from the uncoupled matter pressure $p$ in as much as it depends also on the scalar field through the interacting term.
For example, for a pressure-less fluid, where $p=0$, in general $\tilde{p}\neq 0$. Note also that for the derivative coupling considered here, the energy density of the coupled and uncoupled matter fluids coincide, namely $\rho=\tilde\rho$ since $\rho_{\rm int}=0$.

For the covariant conservations of the ``bare'' $T_{\mu\nu}$ one instead finds
\begin{align}
  U^\nu \nabla^\mu {T}_{\mu\nu} = 0 \qquad\mbox{and}\qquad h_\mu^\nu\nabla^\lambda {T}_{\nu\lambda} = 2n U^\lambda \nabla_{[\lambda} \left[ \left( n \frac{\partial f}{\partial n} U_{\mu]} U^\nu - f \delta_{\mu]}^\nu \right) \nabla_\nu\phi \right] \,,
    \label{054b}
\end{align}
where square brackets denotes anti-symmetrisation. This implies in particular that the ``bare'' matter energy-momentum tensor is not conserved in the presence of the coupling with the scalar field
\begin{align}
  \nabla_\mu T^{\mu\nu} \neq 0 \,.
\end{align}
This is not unexpected since the Einstein field equations imply the conservation of the total energy-momentum tensor only. Any additional conservation equations need to be imposed separately.

\subsection{Derivation of the conservation equations}

The derivations of Eqs.~(\ref{053b}) and (\ref{054b}) are slightly involved and we show some details here. These calculations are similar to the ones appearing in the Appendix of \cite{part1}. Recalling that $U^\mu\nabla_\nu U_\mu=0$ and using Eqs.~(\ref{019b}), we have
\begin{align}
  U^\lambda \nabla^\mu \tilde{T}_{\mu\nu} &= U^\lambda \nabla^\mu \left( p g_{\mu\lambda} + (\rho+p) U_\mu U_\lambda -n^2 \frac{\partial f}{\partial n} U^\alpha \partial_\alpha\phi h_{\mu\lambda} \right) \\
  &= -U_\mu\nabla^\mu\rho -(\rho+p) \nabla^\mu U_\mu -n^2 \frac{\partial f}{\partial n} U^\alpha \partial_\alpha\phi U^\lambda \nabla^\mu h_{\mu\lambda} \\
  &= -\frac{\partial\rho}{\partial n} U_\mu \nabla^\mu n -n \frac{\partial\rho}{\partial n} \nabla^\mu U_\mu -n^2 \frac{\partial f}{\partial n} U^\alpha \partial_\alpha\phi U^\lambda \nabla^\mu \left( U_\lambda U_\mu \right) \\
  &= -\frac{\partial\rho}{\partial n} \nabla^\mu \left( U_\mu n \right) + n^2 \frac{\partial f}{\partial n} U^\alpha \nabla_\alpha\phi \nabla^\mu U_\mu \\
  &= U^\lambda \left( n^2 \frac{\partial\rho}{\partial n} \nabla^\mu U_\mu \right) \nabla_\lambda\phi \,,
\end{align}
in agreement with the first of Eqs.~(\ref{053b}).
Note that if we consider $T_{\mu\nu}$ instead of $\tilde{T}_{\mu\nu}$ in the passages above, we obtain
\begin{align}
  U^\lambda \nabla^\mu {T}_{\mu\nu} = 0 \,,
\end{align}
proving the first of Eqs.~(\ref{054b}).
For the second of Eqs.~(\ref{053b}) we have
\begin{align}
  h_\mu^\nu\nabla^\lambda \tilde{T}_{\nu\lambda} = h_\mu^\nu\nabla^\lambda {T}_{\nu\lambda} - h^\nu_\mu \nabla^\lambda \left( n^2 \frac{\partial f}{\partial n} U^\alpha\partial_\alpha\phi h_{\lambda\nu} \right) \,. \label{app:015}
\end{align}
For the first term we can use the relation (see the appendix of \cite{part1})
\begin{align}
  h_{\mu\nu}\nabla_\lambda {T}^{\nu\lambda} = 2 n U^\lambda\nabla_{[\lambda}(\mu U_{\mu]}) \,,
  \label{app:011}
\end{align}
which gives
\begin{align}
  h_\mu^\nu\nabla^\lambda \tilde{T}_{\nu\lambda} = 2 n U^\lambda\nabla_{[\lambda}(\mu U_{\mu]})- h^\nu_\mu \nabla^\lambda \left( n^2 \frac{\partial f}{\partial n} U^\alpha\partial_\alpha\phi h_{\lambda\nu} \right) \,.
\end{align}
Then making use of Eq.~(\ref{046b}) we obtain
\begin{align}
  h_\mu^\nu\nabla^\lambda \tilde{T}_{\nu\lambda} &= -2 n U^\lambda \left[ \nabla_{[\lambda}\nabla_{\mu]}\varphi + s \nabla_{[\lambda}\nabla_{\mu]}\theta + \nabla_{[\lambda}\left(\beta_A\nabla_{\mu]}\alpha^A\right) \right] +2n U^\lambda \nabla_{[\lambda} \left[ \left( n \frac{\partial f}{\partial n} U_{\mu]} U^\nu - f \delta_{\mu]}^\nu \right) \nabla_\nu\phi \right] \nonumber\\ &\qquad - h^\nu_\mu \nabla^\lambda \left( n^2 \frac{\partial f}{\partial n} U^\alpha\partial_\alpha\phi h_{\lambda\nu} \right) \label{app:012} \\
    &= n U^\lambda \nabla_{\lambda} \left[ \left( n \frac{\partial f}{\partial n} U_{\mu} U^\nu - f \delta_{\mu}^\nu \right) \nabla_\nu\phi \right] - n U^\lambda \nabla_{\mu} \left[ \left( n \frac{\partial f}{\partial n} U_{\lambda} U^\nu - f \delta_{\lambda}^\nu \right) \nabla_\nu\phi \right] - h^\nu_\mu \nabla^\lambda \left( n^2 \frac{\partial f}{\partial n} U^\alpha\partial_\alpha\phi h_{\lambda\nu} \right) \label{app:013} \\
    &= n^2\frac{\partial f}{\partial n} \nabla_\lambda U^\lambda \left( \nabla_\mu\phi + U^\nu U_\mu \nabla_\nu\phi \right) \label{app:014} \\
    &= h^\nu_\mu \left( n^2 \frac{\partial f}{\partial n} \nabla_\lambda U^\lambda \right) \nabla_\nu\phi \,, \label{997}
\end{align}
where to go from line (\ref{app:012}) to (\ref{app:013}) we used the fact that covariant derivatives commute on any scalar and we applied Eqs.~(\ref{004b})-(\ref{007b}); while the passage from line (\ref{app:013}) to (\ref{app:014}) requires long but standard algebraic manipulations together with the use of Eqs.~(\ref{019b}).
This proves the second of Eqs.~(\ref{053b}). Note that if the same calculations (\ref{app:015})--(\ref{app:012}) are applied to $T_{\mu\nu}$ one immediately finds
\begin{align}
  h_\mu^\nu\nabla^\lambda {T}_{\nu\lambda} = 2n U^\lambda \nabla_{[\lambda} \left[ \left( n \frac{\partial f}{\partial n} U_{\mu]} U^\nu - f \delta_{\mu]}^\nu \right) \nabla_\nu\phi \right] \,,
  \label{998}
\end{align}
in agreement with the second of Eqs.~(\ref{054b}).

\section{Cosmology}
\label{sec:cosmology}

In this section we study cosmological applications of the theoretical framework outlined in the previous section. We will first derive the background equations and then rewrite them in the form of a dynamical system. Subsequently, employing dynamical systems techniques, we will study the universe evolution given by two particular interacting models defined within our approach.

\subsection{Cosmological equations}

We will now assume a flat Friedmann-Robertson-Walker (FRW) metric
\begin{align}
ds^2 = -dt^2 + a(t)^2 \left(dx^2+dy^2+dz^2\right) \,,
\end{align}
where $a(t)$ is the cosmological scale factor and $(t,x,y,z)$ are comoving Cartesian coordinates. We will also assume that all the dynamical quantities are homogeneous, i.e.~they depend only on the cosmological time $t$. In particular we will have that $\phi$, $\rho$, $n$, $s$ will be functions of $t$ only. Moreover taking into account comoving coordinates the perfect fluid 4-velocity becomes simply $U^\mu=(-1,0,0,0)$.

We can then derive the following cosmological equations from the Einstein equations (\ref{043b})
\begin{align}
\frac{3H^2}{\kappa^2} &= \rho +\frac{1}{2}\dot\phi^2 +V \,,\label{029b}\\
-\frac{1}{\kappa^2}\left(2\dot H+3H^2\right) &= p+\frac{1}{2}\dot\phi^2 -V -n^2\frac{\partial f}{\partial n}\dot\phi \,, \label{eq:acc}
\end{align}
where $H=\dot a/a$ is the Hubble rate and the over-dot denotes differentiation with respect to $t$. The scalar field equation is instead
\begin{align}
\ddot\phi + 3H\dot\phi +V' -n^2\frac{\partial f}{\partial n} 3H =0 \,,
\label{031b}
\end{align}
and is obtained from (\ref{044b}). We will work with the standard exponential potential $V = V_0 \exp(-\lambda\kappa\phi)$, leaving the analysis for different potentials to future work. We can immediately notice that the Friedmann equation (\ref{029b}) is not modified by the interacting term. This happens because the time-time component of (\ref{030b}) vanishes for the background FRW metric, or equivalently because $T_{\mu\nu}^{\rm (int)}$ is orthogonal to the fluid flow as pointed out for Eq.~(\ref{030b}).

In a cosmological framework we will always have that Eqs.~(\ref{019b}) give the constraints
\begin{align}
  \dot n +3H n =0 \quad\mbox{and}\quad \dot s=0 \,.
  \label{eqn:bg0}
\end{align}
These equations tells us that the entropy density per particle is conserved through the universe's evolution, while the particle density decays according to
\begin{align}
n\propto a^{-3} \,,
\end{align}
which is expected from geometrical considerations. The two dynamical quantities ultimately appearing in Eqs.~(\ref{029b})--(\ref{031b}) are thus $\phi(t)$ and $a(t)$ only.

\subsection{Cosmological dynamics}

In this section we will examine the dynamics of a universe described by the cosmological equations~(\ref{029b})--(\ref{031b}) using dynamical systems techniques. We begin by introducing the canonical dimensionless variables \cite{Copeland:1997et,Tamanini:2014mpa}
\begin{align}
\sigma=\frac{\kappa \sqrt{\rho}}{\sqrt{3}H}\,, \quad   x=\frac{\kappa \dot\phi}{\sqrt{6}H}\,, \quad y=\frac{\kappa \sqrt{V}}{\sqrt{3}H} \,.
\end{align}
which give us the Friedmann constraint
\begin{align}
1=\sigma^2+x^2+y^2 \,.
\end{align}
Using these we can rewrite the cosmological equations as the dynamical system 
\begin{align}
  x'&=-\frac{1}{2}(3x((w-1)x^2+(w+1)y^2+1-w)-\sqrt{6}(A(x^2-1)+\lambda y^2))\label{100} \,,\\
  y'&=-\frac{1}{2}y(3((w-1)x^2+(w+1)(y^2-1))+\sqrt{6}x(\lambda-A)) \,. \label{101}
\end{align}
The quantity $A$ is defined by
\begin{align}
A=-\frac{\kappa}{H}n^2\frac{\partial f}{\partial n} \,.
\end{align}
Hence the dynamical system is not closed until one specifies the form of the function $f$.  For particular choices of $f$ we can rewrite $A$ as a function of the two dimensionless variables $x$ and $y$. These special models do not increase the dimensions of the dynamical system which remains 2-dimensional. In general, however, if $A$ cannot be written in terms of $x$ and $y$ only, the dimension of the phase space will increase. This does not hinder the use of dynamical system techniques, however, visualising the phase space might become more involved. An example is a coupling of the form $f=-f_0\exp(\gamma\kappa\phi)/n$ with $\gamma$ a dimensionless constant. In this case $A=-\kappa f_0 \exp(\gamma\kappa\phi)/H$ which cannot be expressed in terms of $x$ and $y$ only.

In the following we will consider one 2 dimensional and one 3 dimensional model, with two choices for $f$ as displayed in Tab.~\ref{tabf}, where $\alpha$, $\gamma$ and $\xi$ are dimensionless constants, with $\gamma$ algebraically related to $\xi$ by
\begin{align}
\gamma=\frac{\xi}{(\frac{1}{2}-\alpha)(w+1)-1} \,.
\end{align}
\begin{table}[!ht]
\begin{tabular}{c|c|c}
\mbox{} & $f$ & $A$ \\
\hline
Model A & $ -\gamma \rho^{1/2-\alpha}V^{\alpha}/\sqrt{3}n$ & $\xi y^{2\alpha}(1-x^2-y^2)^{1/2-\alpha}$ \\
Model B & $\frac{\xi H_0}{\kappa n}$  & $\xi\frac{H_0}{H}$
\end{tabular}
\caption{Choices of the interacting function for the dynamical systems analysis.}
\label{tabf}
\end{table}

These models are studied using the standard approach of dynamical systems analysis. The critical points of the respective models will be found, their corresponding eigenvalues will be computed, and these results will be interpreted within the context of cosmology.
When possible the phase space of the dynamical system will be drawn together with few trajectories numerically computed for particular choices of the model parameters.

\subsubsection{Model A: 2D dynamical system}

In this subsection we analyse model A of the dynamical system~(\ref{100})-(\ref{101}). This corresponds to choosing the function $f$ to be
\begin{align}
f\propto\frac{\rho^{1/2-\alpha}V^{\alpha}}{\sqrt{3}n} \,. 
\end{align}
In this case the dynamical system remains two dimensional, as $A$ can be written in terms of $x$ and $y$ as
\begin{align}
A=\xi y^{2\alpha}(1-x^2-y^2)^{1/2-\alpha}
\end{align}
with $\xi$ a constant parameter. The Friedmann equations can be rewritten to give the acceleration equation
\begin{align}
\frac{\dot{H}}{H^2}=\frac{3}{2}[-(1+w)+(w-1)x^2+(1+w)y^2-\sqrt{\frac{2}{3}}\xi x y^{2\alpha}(1-x^2-y^2)^{\frac{1}{2}-\alpha}]
\end{align}
which means we can define the effective equation of state (EoS)
\begin{align}
w_{\rm eff}= x^2-y^2+w(1-x^2-y^2)+\sqrt{\frac{2}{3}}\xi x y^{2\alpha}(1-x^2-y^2)^{\frac{1}{2}-\alpha}.
\end{align}
Critical points correspond to solving the system
\begin{align}
x'=0, \quad y'=0.
\end{align}
For general values of $\alpha$ finding critical points is difficult, and the system becomes singular unless $\alpha$ lies in the range $0\leq \alpha \leq 1/2$, so in the following we will examine some specific choices of $\alpha$ lying in this range. The simplest such choice is $\alpha=1/2$. This corresponds to choosing $f=\xi \sqrt{V}/(3n)$, meaning $A=\xi y$. The background dynamics of this model corresponds to the one of a particular $k$-essence scalar field analysed in \cite{Tamanini:2014mpa}. The phenomenology at cosmological distances is quite interesting and includes late time phantom dominated solutions with dynamical crossing of the phantom barrier. Moreover while the model considered in \cite{Tamanini:2014mpa} always exhibits instabilities at the level of cosmological perturbations, the one constructed with the formalism of this work could be stable since the dynamics at the level of perturbations will be completely different.
A more in depth discussion on such model, namely for the choice $f=\xi \sqrt{V}/(3n)$, will be given in Sec.~\ref{sec:conclusion}.

In the remainder of this section we will examine the case $\alpha=0$. We set $w=0$ for simplicity, since other values of $w$ are not relevant for dark matter interacting models.  The critical points of the dynamical system are displayed in Tab.~\ref{critA}.
\begin{table}[!ht]
\begin{tabular}{|c|c|c|}
  \hline
  Point & $x$ & $y$  \\
  \hline
  \hline
  $A_{\pm}$ & $\pm 1$ & 0 \\
  \hline
  $C$ & $\frac{\lambda}{\sqrt{6}}$ & $\sqrt{1-\frac{\lambda^2}{6}}$  \\
  \hline
  $D$ & $-\frac{\sqrt{2}\xi}{\sqrt{3+2\xi^2}}$ &
  0  \\
  \hline
\end{tabular}
\caption{Critical points of Model A with $w=0$ and $\alpha=0$.}
\label{critA}
\end{table}

Properties of these critical points, including existence and stability can be found in Table~\ref{tab05}. The system has potentially up to four critical points depending on the values of the parameters $\lambda$ and $\xi$:
\begin{itemize}
\item {\it Point $A_{\pm}$}. These two points exist for all $\lambda$ and $\xi$. They are the standard solutions dominated by the scalar field kinetic energy, with effective equation of state of a stiff fluid $w_{\rm eff}=1$. These points are either unstable or saddle points depending on whether the absolute value of $\lambda$ is less than $\sqrt{6}$. 
\item {\it Point C.} This point corresponds to a universe completely dominated by a scalar field. It exists only for $\lambda^2<6$. It is stable when  $\xi^2<2(3-\lambda^2)$, and a saddle node otherwise. The point describes an accelerating universe when the scalar potential is sufficiently flat, requiring $\lambda^2<2$.
\item {\it Point D.} In the limit $\xi\rightarrow0$ this point reduces to the origin. The energy density of this point is dominated both by the matter and the kinetic energy of the scalar field, with no scalar field potential energy density contribution. It is the stable late time attractor only for negative $\xi$ satisfying $\lambda \xi< -\frac{2}{3}\sqrt{1+\frac{2\xi^2}{3}}$. The effective EoS is $w_{\rm eff}=0$, the same as a matter dominated solution, however it should be noted that for general $w$ it is not a scaling solution, generally $w_{\rm eff}=3w/(3+2\xi^2)$.
\end{itemize}

\begin{table}[!ht]
\begin{tabular}{|c|c|c|c|c|}
\hline
\mbox{Point} & Existence & $w_{\rm eff}$ & Acceleration & Stability \\
\hline \hline
$A_{-}$ & $\forall \lambda,\xi$ & 1 & No & Unstable node: $ \lambda>-\sqrt{6}$ \\
& & & & Saddle node: otherwise \\
\hline
$A_{+}$ & $\forall \lambda,\xi$ & 1 & No & Unstable node: $ \lambda<\sqrt{6}$ \\
& & & & Saddle node: otherwise 
\\ \hline
$C$ & $\lambda^2<6$ & $\frac{\lambda^2-3}{3}$ & $\lambda^2<2$ & Stable node: $\xi^2<2(3-\lambda^2)$ \\
& & & & Saddle point: $\xi^2>2(3-\lambda^2)$ \\
\hline
$D$ & $\forall \lambda,\xi$ &  0 & No  & Stable node: $\lambda\xi<-\frac{3}{2}\sqrt{1+\frac{2\xi^2}{3}}$ \\
& & & & Saddle node: otherwise
\\
\hline
\end{tabular}
\caption{Stability of critical points of Model A with $w=0$ and $\alpha=0$} \label{tab05} 
\end{table}

\begin{figure}[!ht]
\includegraphics[width=0.48\textwidth]{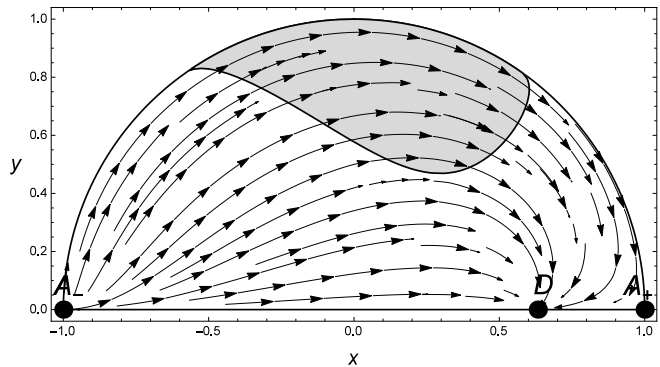}
\caption{Phase space for the dynamical system when $\xi=-1$ and $\lambda=4$. The shaded region indicates where acceleration is present.}\label{xi=-1} 
\end{figure}

\begin{figure}[!ht]
\includegraphics[width=0.48\textwidth]{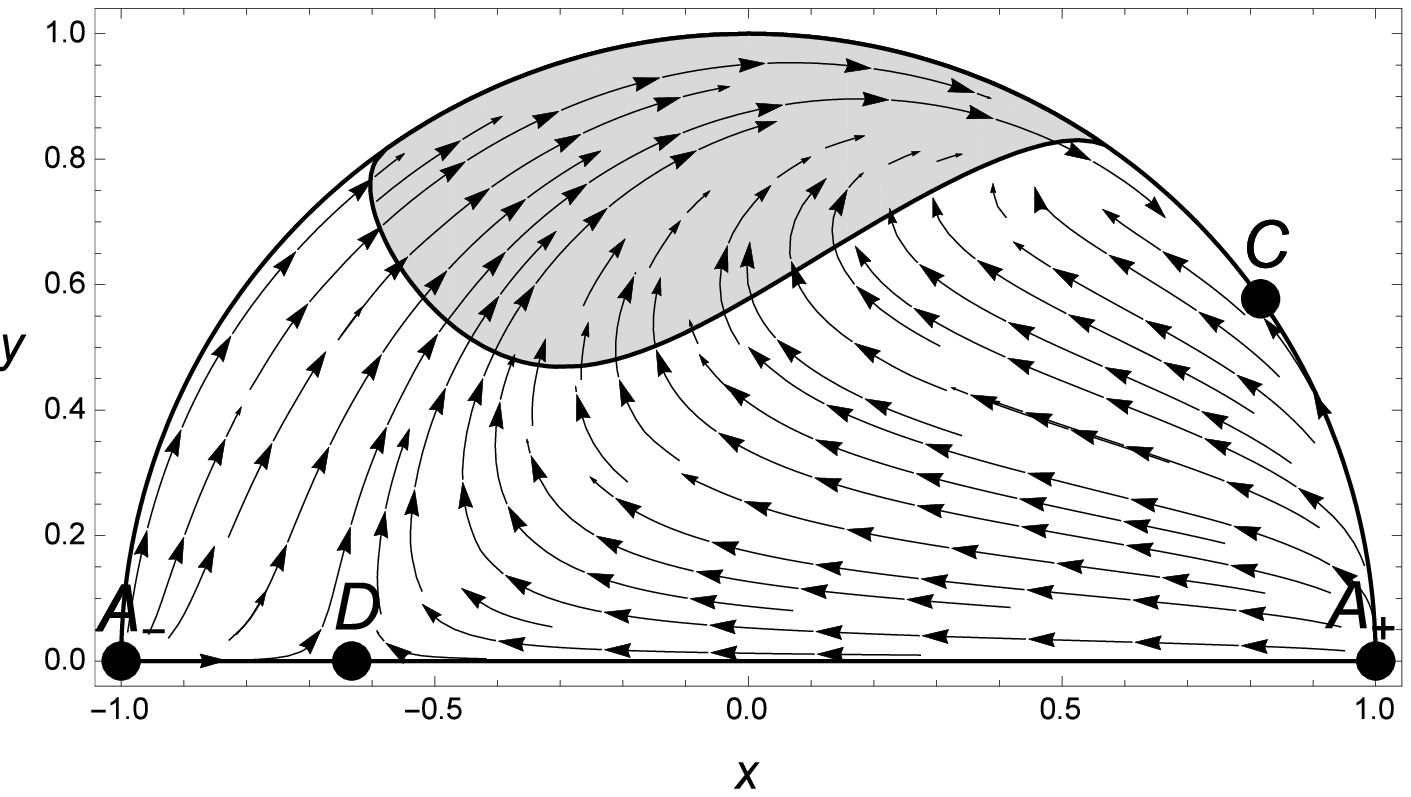}
\caption{Phase space for the dynamical system when $\xi=1$ and $\lambda=2$. The shaded region indicates where acceleration is present.}\label{xi=1lambda=2}
\end{figure}

\begin{figure}[!ht]
\includegraphics[width=0.48\textwidth]{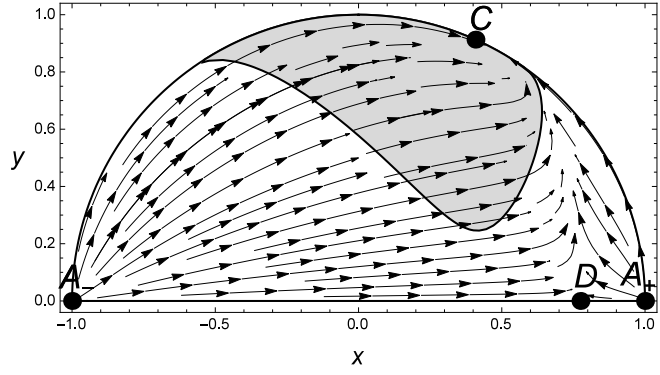}
\caption{Phase space for the dynamical system when $\xi=-1.5$ and $\lambda=1$. The shaded region indicates where acceleration is present.}\label{xi=-3/2lambda=1}
\end{figure}

The phase space for this two dimensional model is simply the upper half unit disc. We show the phase space diagrams with example trajectories in Fig.~\ref{xi=-1}, Fig.~\ref{xi=1lambda=2} and Fig.~\ref{xi=-3/2lambda=1}. The region of acceleration is indicated by the grey region. The shape of this region of acceleration is dependent only on $\xi$, and is independent of the parameter $\lambda$.  In Fig.~\ref{xi=-1} we make the parameter choice $\lambda=4$ and $\xi=-1$. Trajectories begin at the stiff matter point $A_-$. Many trajectories then undergo a short transient accelerating phase, before then decelerating, with some trajectories being drawn towards the saddle point $A_+$. Finally the late time attractor $D$ is reached, which has a matter equation of state.

In Fig.~\ref{xi=1lambda=2} we choose $\lambda=2$ and $\xi=+1$. Trajectories now start at either of the stiff matter points $A_+$ or $A_-$.  Trajectories are then drawn towards the saddle matter dominated point $D$.  Trajectories are then attracted upwards, passing through the region of acceleration. They are then finally attracted towards the global attractor $C$, which is not accelerating for this choice of parameter values. 

In Fig.~\ref{xi=-3/2lambda=1} the parameter values $\lambda=1$ and $\xi=-3/2$ are chosen. Again trajectories begin at the stiff matter points $A_+$ and $A_-$ before passing through the matter dominated saddle point $D$. Trajectories then enter the region of acceleration and end at the late time global attractor $C$, which in this case is accelerating as the scalar potential is sufficiently flat. 

These models are able to accurately describe late time universe phenomenology, there are many parameter choices which result in a global accelerating attractor. They can also describe universes undergoing transient periods of acceleration. However as is typical in these models it breaks down at early times, as it begins at the stiff matter point $A_-$, whose effective EoS $w_{\rm eff}=1$ is not physically viable. 

There are a few key differences with these models if compared to the canonical scalar field case, which we recover by taking the limit $\xi\rightarrow0$.  There is no longer a scaling solution in this model, however the origin $O$ is deformed into the point $D$ which behaves as though the universe is matter dominated. Moreover, unlike the origin $O$ this point $D$ can be stable for a variety of parameter choices.

\subsubsection{Model B: 3D dynamical system}
\label{sec:modelB}

In this section we consider model B of the dynamical system~(\ref{100})-(\ref{101}) where we take the coupling function $f$ to simply be proportional to $1/n$, as outlined in Tab.~\ref{tabf}. In particular we take
\begin{align}
f=\frac{\xi H_0}{\kappa n} .
\label{eq:simple_coupling}
\end{align}
where $\xi$ is a constant and $H_0$ is the Hubble parameter at an arbitrary fixed time. The function $A$ is  now simply
\begin{align}
A=\xi\frac{H_0}{H}
\end{align} 
The resulting dynamical system is no longer two dimensional, in which case we need to introduce a third variable, see for instance \cite{Boehmer:2008av}. We take this to be
\begin{align}
z=\frac{H_0}{H_0+H}
\end{align} 
which is compact and lies in the range $0\leq z<1$. The resulting dynamical system is given by
\begin{align}
  x'&=-\frac{1}{2(z-1)}(3x(z-1)(1-w+(w-1)x^2+(1+w)y^2)+\sqrt{6}(\xi z(1-x^2)-\lambda (z-1) y^2))\label{107}\\
  y'&=-\frac{y}{2(z-1)}(3(z-1)((1+w)(y^2-1)+(w-1)x^2)+\sqrt{6}x((z-1)\lambda+z\xi))\label{108} \\
  z'&=\frac{1}{2}z(3(z-1)((1+w)(y^2-1)+(w-1)x^2)+\sqrt{6}z\xi x). \label{109}
\end{align}
The Friedman equations can again be rearranged to give the acceleration equation in terms of these new variables $x, y$ and $z$. This time for the acceleration equation we find
\begin{align}
\frac{\dot{H}}{H^2}=\frac{3}{2}[-(1+w)-(1-w)x^2+(1+w)y^2-\frac{2\xi}{\sqrt{6}}\frac{zx}{1-z}].
\end{align}
Hence in this model $w_{\rm eff}$ is given by
\begin{align}
w_{\rm eff}=w+(1-w)x^2-(w+1)y^2+\frac{2\xi}{\sqrt{6}}\frac{zx}{1-z}.
\label{999}
\end{align}
The variable $z$ was chosen so that the phase space of the system is compact, this time the phase space will be a semi-cylinder of unit height. The dynamical system has a singularity on the plane $z=1$, which corresponds to the Hubble parameter $H \rightarrow 0$, i.e.~when $a\rightarrow\infty$, usually happening at $t\rightarrow\infty$ . However, none of the critical points of the system lie on this plane. We display the critical points for the system in Tab~\ref{critB}.
\begin{table}[!ht]
\begin{tabular}{|c|c|c|c|}
  \hline
  Point & $x$ & $y$ & $z$ \\
  \hline
  \hline
  $O$ & 0 & 0 & 0 \\
  \hline
  $A_{\pm}$ & $\pm 1$ & 0 & 0 \\
  \hline
  $B$ & $\sqrt{\frac{3}{2}}  \frac{(1+w)}{\lambda} $ & 
 $ \sqrt{\frac{3}{2}} \frac{\sqrt{(1+w)(1-w)}}{\lambda}$ & 0 \\
  \hline
  $C$ & $\frac{\lambda}{\sqrt{6}}$ & $\sqrt{1-\frac{\lambda^2}{6}}$ & 0 \\
  \hline
  $D$ & 0 & 1 & $ \frac{\lambda}{\lambda+\xi}$ \\
  \hline
  $E_{\pm}$ & $\pm 1$ & 0 &   $\frac{\sqrt{6}}{\sqrt{6}\mp \xi}$ \\
  \hline
\end{tabular}
\caption{Critical points of Model B}
\label{critB}
\end{table}

The stability of the critical points is shown in Tab.~\ref{tab06}. The system has at most seven critical points at any one time. The points $O$, $A_{\pm}$, $B$ and $C$ lie on the $z=0$ plane and have exactly the same coordinates as the canonical scalar field \cite{Copeland:1997et}. Moreover, the existence and acceleration properties of these points remain the same as the canonical case. This happens exactly because the contribution of the interaction vanishes for $z=0$, as can be realised looking at Eqs.~(\ref{107})--(\ref{999}). However, none of these points can now be stable, the $z=0$ plane is unstable in general and trajectories starting on this plane will not stay there. There are three new critical points in this model, given by:
\begin{itemize} 
\item {\it Point $D$} This point exists only for $\xi>0$. At this point the energy density of the universe is dominated by the scalar field potential energy, with zero matter and scalar field kinetic energy contributions. When this point exists,  it is always the late time global attractor. Moreover it has an effective EoS $w_{\rm eff}=-1$, hence acceleration is present and the behaviour of the cosmological constant is mimicked. 
\item {\it Point $E_-$} This point only exists for $\xi>0$. The energy density of the universe at this point is dominated by the kinetic energy of the scalar field, with no potential or matter contributions. When this point exists it is the unique late time attractor. It also has the effective EoS $w_{\rm eff}=-1$.
\item {\it Point $E_+$} This point is similar to $E_-$, except it only exists for $\xi<0$. It is also dominated by the scalar field kinetic energy, and has effective EoS $w_{\rm eff}=-1$. However unlike $E_-$, this point is always a saddle point and is unstable in general. 
\end{itemize}

\begin{table}[!ht]
\begin{tabular}{|c|c|c|c|c|}
\hline
\mbox{Point} & Existence & $w_{\rm eff}$ & Acceleration & Stability \\
\hline \hline
$O$ &  $\forall \lambda,\xi$ & $w$ & No & Saddle node \\ \hline
$A_{-}$ & $\forall \lambda,\xi$ & 1 & No & Unstable node: $ \lambda>-\sqrt{6}$ \\
& & & & Saddle node: otherwise \\
\hline
$A_{+}$ & $\forall \lambda,\xi$ & 1 & No & Unstable node: $ \lambda<\sqrt{6}$ \\
& & & & Saddle node: otherwise \\
\hline
$B$ & $\lambda^2>3(w+1)$ & $w$ & No & Saddle point
\\
\hline
$C$ & $\lambda^2<6$ & $\frac{\lambda^2-3}{3}$ & $\lambda^2<2$ & Saddle point \\
\hline
$D$ & $\xi>0$ & $-1$ & Yes & Stable node: $\lambda^2<3/2$ \\
& & & & Stable spiral: $\lambda^2>3/2$  \\ \hline
$E_{+} $& $\xi<0$ & $-1$ & Yes & Stable node 
\\
\hline
$E_{-}$ & $\xi>0 $  & $-1$ & Yes & Saddle point
\\
\hline
\end{tabular}
\caption{Stability of critical points of Model B} \label{tab06}
\end{table}

The global dynamics of these models are particularly simple to analyse. The late time global attractor is either $D$, in the case $\xi>0$, or $E_+$, in the case $\xi<0$. Moreover both of these late time global attractors have effective EoS $w_{\rm eff}=-1$.  Trajectories begin on the $z=0$ plane, and while they remain on this plane the trajectories behave as in the case of the canonical scalar field. This plane is generically unstable and the trajectories eventually leave this plane and arrive at the late time attractor. 

\begin{figure}[!ht]
\includegraphics[width=0.70\textwidth]{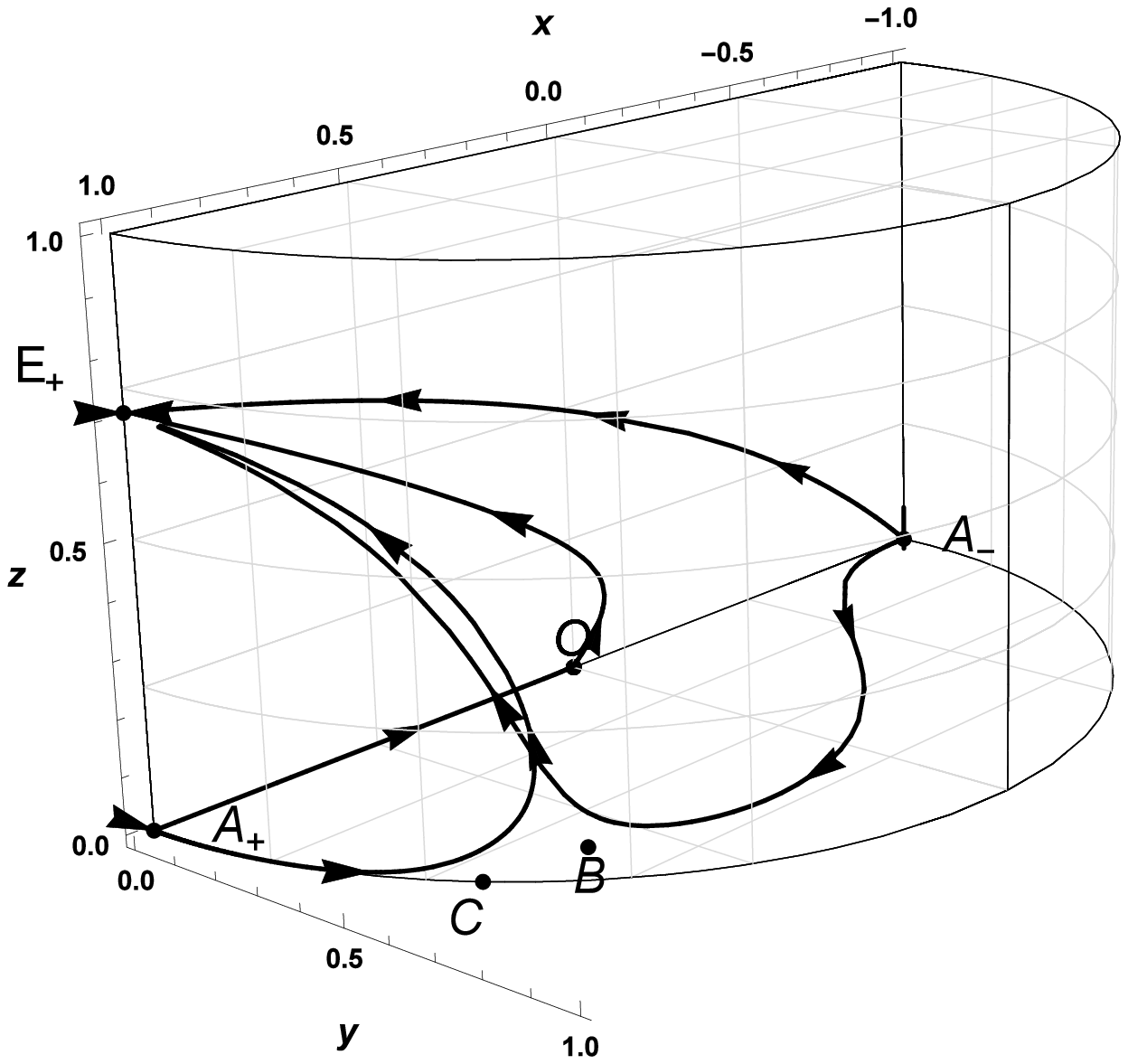}
\caption{Phase space of model B of the dynamical system with values $w=0$, $\lambda=2$, $\xi=-1$.} \label{3dxi-1}
\end{figure}

\begin{figure}[!ht]
\includegraphics[width=0.70\textwidth]{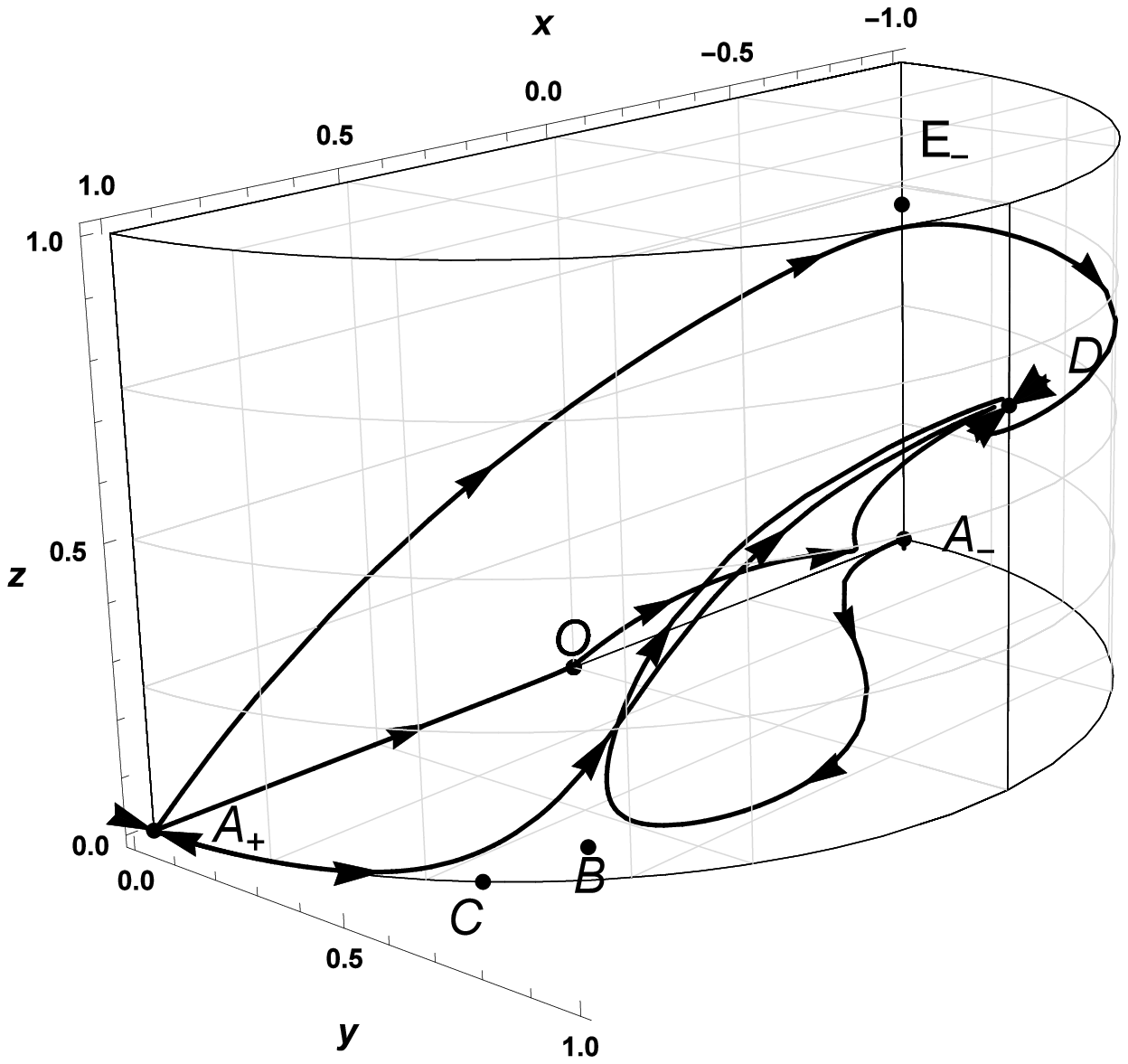}
\caption{Phase space of model B of the dynamical system with values $w=0$, $\lambda=2$, $\xi=1$.}
\label{3dxi1}
\end{figure}

Phase spaces for a couple of parameter choices with example trajectories are plotted in Fig.~\ref{3dxi-1} and Fig.~\ref{3dxi1}. In Fig.~\ref{3dxi-1}, the parameter values $\lambda=2$, $\xi=-1$ are chosen. Trajectories start at the stiff matter points $A_{\pm}$, before being attracted towards one of the three saddles $O$, $B$ or $C$ on the $z=0$ plane. All trajectories then end at the late time accelerating global attractor $E_+$.  There is a particular (heteroclinic) trajectory of interest that passes through the matter dominated point $O$ before ending at the accelerating point $E_+$. Ignoring the theoretical issues at early times, this trajectory would mirror the dynamics of a universe with no scalar field potential but still with a cosmological constant late time behaviour.

In Fig.~\ref{3dxi1} the parameter values are set to be $\lambda=2$ and $\xi=+1$. Trajectories start at the stiff matter points $A_{\pm}$, and are then attracted towards the saddle points on the plane $z=0$. The trajectories then leave the plane, with some attracted towards the saddle point $E_-$. All trajectories end up at the late time accelerating attractor $D$. 

These models are able to accurately describe late time universe phenomenology. All trajectories end at accelerating critical points with effective equation of state $w_{\rm eff}=-1$. In this scenario the dynamics of the universe would mirror that of a universe with no scalar field but with a cosmological constant and thus these models could represent a solution to the cosmological constant problem.

\subsection{A glimpse at the perturbations}
\label{sec:perturbations}

In this subsection we will briefly consider scalar perturbations of Model~B in the linear approximation. Note that from the dynamical systems analysis of Sec.~\ref{sec:modelB} it is clear that the late time attractor of this model is always represented by a de Sitter solution, which is well suited to describe dark energy domination. Moreover, this is the simplest scalar-fluid coupling where the gradient of the scalar field $\phi$ appears. It will thus be interesting to see how the cosmological perturbation equations are modified in presence of such a coupling. In what follows we will show the perturbed equations in Newtonian gauge and briefly discuss their main features. A complete investigation of the dynamics of cosmological perturbations is outside the scope of the present work, but will be delivered in forthcoming studies~\cite{perturbations}.

We will study the behaviour of scalar perturbations assuming the following metric in Newtonian gauge using Cartesian coordinates
\begin{align}
  ds^2 = - (1+2\Phi) dt^2 + 
  \frac{(1-2\Psi)\, a^2 }{\left[1+\frac{1}{4} k (x^2+y^2+z^2)\right]^2} (dx^2+dy^2+dz^2) \,,
\end{align}
where $k=-1,0,1$, and $\Psi$ and $\Phi$ are functions of all the coordinates. Since in both the explicit and implicit frames all the matter sources can be written as perfect fluids, no anisotropies appear in the scalar-fluid models considered here. In fact, the off-diagonal spatial field equations immediately imply that $\Phi = \Psi$, which will be used to simplify the other perturbed equations. We will also give the equations directly in the Fourier space: $\nabla^2 \mapsto -q^2$, where $\nabla^2$ is the Laplace operator.

The covariant field equations to perturb are given by Eqs.~(\ref{043b})--(\ref{044b}). As we mentioned, we will restrict the equations to the particular coupling~(\ref{eq:simple_coupling}), and, in order to simplify the notation, we will also redefine the coupling constant as $\hat\xi = \xi H_0 / \kappa$. Now, $\rho$, $p$ and $\phi$ will refer to background quantities, while $\Psi$, $\delta\phi$, $\delta\rho$ and $\delta p$ will denote small perturbations. Furthermore, the perturbed 4-velocity of the fluid is
\begin{align}
  \delta U_\mu = (-\Psi, \partial_i v) \,,
\end{align}
with $v$ being the scalar perturbation of the matter fluid velocity. The time component of this relation, with $\Psi$, follows from the constraint $U_\mu U^\mu = -1$.

At this point we are ready to state the equations for the cosmological perturbations. The $00$-component of the perturbed Einstein field equations then reads
\begin{align}
  \left(\frac{q^2}{a^2}+8 \pi  \rho+8 \pi  V -6\frac{k}{a^2}\right) \Psi
  +3 H \dot\Psi +4 \pi  V' \delta\phi+4 \pi  \delta\rho+4 \pi  \dot\phi \dot{\delta\phi} =0 \,,
  \label{eq:023}
\end{align}
while the $0i$-components are
\begin{align}
  4 \pi  \left(\hat\xi \dot\phi+p+\rho\right) v
  -4 \pi  \dot\phi \delta\phi + H \Psi +\dot\Psi  = 0 \,,
\end{align}
and the diagonal $ij$-components become
\begin{align}
  \ddot\Psi +4 H \dot\Psi +\left(4 \pi  \hat\xi \dot\phi+2 \dot{H}+3 H^2+4 \pi  \dot\phi^2 -\frac{k}{a^2}\right) \Psi 
  -4 \pi  \left(\hat\xi+\dot\phi\right) \dot{\delta\phi} +4 \pi  V' \delta\phi-4 \pi \delta{p} =0 \,.
\end{align}
The perturbation of the scalar field equation is given by
\begin{align}
  \ddot{\delta\phi} +3 H \dot{\delta\phi} + \left(\frac{q^2}{a^2}+V''\right) \delta\phi
  -\hat\xi \frac{q^2}{a^2}v  +\left(3 H \hat\xi+2 V'\right) \Psi -\left(3 \hat\xi+4 \dot\phi\right) \dot\Psi = 0 \,.
\end{align}
Finally we also provide the perturbations of the matter conservation equations: the time-component is
\begin{align}
  (p+\rho) \frac{q^2}{a^2} v -3 H \delta{p}-3 H \delta\rho+3 (p+\rho) \dot\Psi-\dot{\delta\rho} =0 \,,
  \label{eq:025}
\end{align}
while the spatial components are 
\begin{align}
  \left[\hat\xi \left(\ddot\phi+3 H \dot\phi\right) - 3H c_s^2 \left( p+ \rho\right)\right] v
  +\dot{v} \left(\hat\xi \dot\phi+p+\rho\right)
  +3 \hat\xi H \delta\phi+\hat\xi \dot{\delta\phi}+(p+\rho) \Psi+\delta{p} =0 \,,
  \label{eq:024}
\end{align}
where $c_s^2 = \partial p/\partial\rho$.

We should firstly notice that the coupling does not appear in Eqs.~(\ref{eq:023}) and~(\ref{eq:025}). This is a general feature of these derivative scalar-fluid models as will be shown in~\cite{perturbations}. Mathematically this is due to the fact that the interacting energy-momentum tensor is orthogonal to the fluid's 4-velocity, namely $U^\mu T^{\rm (int)}_{\mu\nu}=0$, as discussed after Eq.~(\ref{eq:111}). In fact, in a cosmological context this implies that the time-component of both, the Einstein and matter conservation equations, reduce to their uncoupled counterparts, as already noticed for the background equations.

The remaining perturbation equations are modified by terms containing $\hat\xi$, which is the only coupling parameter appearing in these models and thus fully characterizing the interaction between dark energy and matter. In the limit $\hat\xi \to 0$ Eqs.~(\ref{eq:023})--(\ref{eq:024}) reduce to the uncoupled correspondent perturbation equations. Note also how the interacting terms mix the gradient of the scalar field and the velocity of the fluid.

This is expected from the derivative coupling used in the Lagrangian, which thus represents interesting new interactions even at the level of perturbations. In~\cite{perturbations} it will be shown that the evolution of the perturbations governed by Eqs.~(\ref{eq:023})--(\ref{eq:024}) are indistinguishable from the $\Lambda$CDM dynamics at sub-horizon scales ($q \gg H^2a^2$), although signatures of the interaction should arise at larger or non-linear scales.

\section{Discussion and conclusion}
\label{sec:conclusion}

The main motivation of this paper was the continuation of the approach outlined in Part~I by taking into account derivative couplings between a scalar field and a barotropic fluid based on a variational formulation. In particular, we were interested in couplings linear in the first partial derivatives of the scalar field. The general expression for such a coupling can be written in the form $f(n,s,\phi) J^\mu\partial_\mu\phi$ and can only be treated with the methods previously developed in Part~I. The presence of the term $J^\mu\partial_\mu\phi$ requires a separate treatment from that in Part~I since the effects given by the appearance of the scalar field's derivative will in general produce a different phenomenology. 

As an example we recall the fact that the energy-momentum (\ref{030b}), arising from the interacting contribution, is always orthogonal to the fluid 4-velocity, namely $U^\mu T_{\mu\nu}^{\rm (int)}=0$. Formally this happens because $T_{\mu\nu}^{\rm (int)}\propto h_{\mu\nu}$ which in turn is due to the fact that in the derivative interacting term (\ref{302b}) the metric tensor $g_{\mu\nu}$ appears within $n$ only. As a consequence $T_{\mu\nu}^{\rm (int)}$ vanishes whenever $f(n,s,\phi)$ does not depend on $n$, as one can easily realize looking at Eq.~(\ref{030b}). The property $U^\mu T_{\mu\nu}^{\rm (int)}=0$ is particularly relevant for cosmological applications since it implies that at the background level the Friedmann equation is never modified by the interaction between the fluid and the scalar field, see Eq.~(\ref{029b}). This is in contrast to the algebraic couplings considered in Part~I where this always happens. Note however that both the acceleration equation (\ref{eq:acc}) and scalar field equation (\ref{031b}) do get modified by the interaction. This implies that the background cosmological evolution can actually differ from the corresponding non-interacting one, though the Friedmann equation forces a matter-like evolution whenever the scalar field energy density becomes negligible if compared to the matter energy density, irrespectively of the strength of the interaction. The fact that $U^\mu T_{\mu\nu}^{\rm (int)}=0$ implies also that the perturbed Einstein equation (\ref{eq:023}) as well as the continuity equation (\ref{eq:025}) are not modified by the interacting term, although in all the other perturbed equations deviations due to the coupling appear.

In Sec.~\ref{sec:cosmology} we have studied some of the cosmological consequences obtained from these models, employing in particular dynamical systems techniques. One would expect that a derivative coupling would complicate the subsequent dynamical systems formulation. However, it turns out that the derivative models studied here are somewhat easier to handle than models based on the algebraic coupling $f(n,s,\phi)$ considered in Part~I. In particular we have considered two types of derivative couplings, labelling them as Models~A and B according to Tab.~\ref{tabf}. The background cosmological evolution of Model~A has been characterized by a 2D dynamical system, while the one of Model~B required the analysis of a 3D dynamical system. For both models we found late time accelerating attractors, capable of describing the present dark energy dominated epoch, and possible intermediate phases of dark matter domination.

For Model~A the specific case corresponding to the parameter $\alpha=0$ has been studied in depth. We found that a dark matter to dark energy transition can be obtained for some values of the model parameters and is always described by the (heteroclinic) trajectories connecting Point~$D$ to Point~$C$ as depicted in Fig.~\ref{xi=-3/2lambda=1}. Note that, as mentioned before, Point~$D$ describes a scaling solution only when $w=0$, i.e.~when the fluid describes (dark) matter. For a different matter EoS one finds $w_{\rm eff}\neq w$, implying that modifications to the standard cosmological evolution might arise if radiation is added into the analysis.

Another simple case of Model~A is achieved by the choice $\alpha=1/2$. As mentioned above, the resulting background cosmological equations corresponding to this choice match the one arising from a particular $k$-essence model studied in \cite{Tamanini:2014mpa}. Since the dynamical analysis of the $\alpha=1/2$ case has already been performed in detail in \cite{Tamanini:2014mpa}, it has not been considered in the present work. However we briefly recall here some of the features and results derived in \cite{Tamanini:2014mpa} which arise as well in the background cosmological dynamics of Model~A with $\alpha=1/2$. The phenomenology found in this case is highly rich with different cosmological behaviours that cannot be obtained with a non-interacting canonical scalar field. At late times it is possible to obtain phantom domination ($w_{\rm eff}<-1$) with dynamical crossing of the phantom barrier, while at early times super-stiff ($w_{\rm eff}>1$) solutions can be attained. Theoretical models predicting a dynamical crossing of the phantom barrier are important to investigate in the eventuality that future surveys will detect (with statistical significance) a dark energy EoS within the phantom regime. Models of such kind obtained within the $k$-essence framework, such as the one considered in \cite{Tamanini:2014mpa}, suffer from instabilities at the level of perturbations \cite{Vikman:2004dc}. Nevertheless if the same phantom crossing evolution is not realized by a $k$-essence model, as in the case of Model~A, then the cosmological perturbations might be stable. This is because, although the background equations coincide, one expects differences in the perturbation equations between Model~A with $\alpha=1/2$ and the model considered in \cite{Tamanini:2014mpa}. These considerations enforce the motivation for analysing the dynamics of cosmological perturbations arising from the quintessence interacting models studied here and in Part~I~\cite{perturbations}.

Considering Model~B, it was found that a global accelerated attractor always appears in the phase space, irrespective of the values of the model parameters. This implies that for this particular coupling late time dark energy domination is always attained. Interestingly one of these attractors (Point~$E_+$), although mimicking a cosmological constant behaviour, is completely dominated by the scalar field's kinetic energy, showing that within Model~B late time acceleration can also be achieved without a self-interacting (and thus massless) scalar field. Moreover for some choices of the parameters a scaling solution attracting all the early time trajectories (the ones on the $z=0$ plane) is present in the phase space; see Point~$B$ in Fig.~\ref{3dxi1}. This saddle point forces the early universe to reach matter domination before switching to the late time accelerated behaviour. Thus for all physically possible initial conditions (for which $z \simeq 0$, i.e.~$H\gg H_0$) a transition from dark energy to dark matter is attained at late times, solving in this manner any fine tuning issue. Although the presence of a scalar field at early times is expected to be strongly constrained by observations \cite{Ade:2015yua}, the results obtained within Model~B are promising and should merit further consideration. For this reason in Sec.~\ref{sec:perturbations} we presented and briefly discussed the scalar cosmological perturbations for such model, although a more detailed investigation will be presented in future works \cite{perturbations}.

\begin{figure}[!tb]
\includegraphics[width=0.48\textwidth]{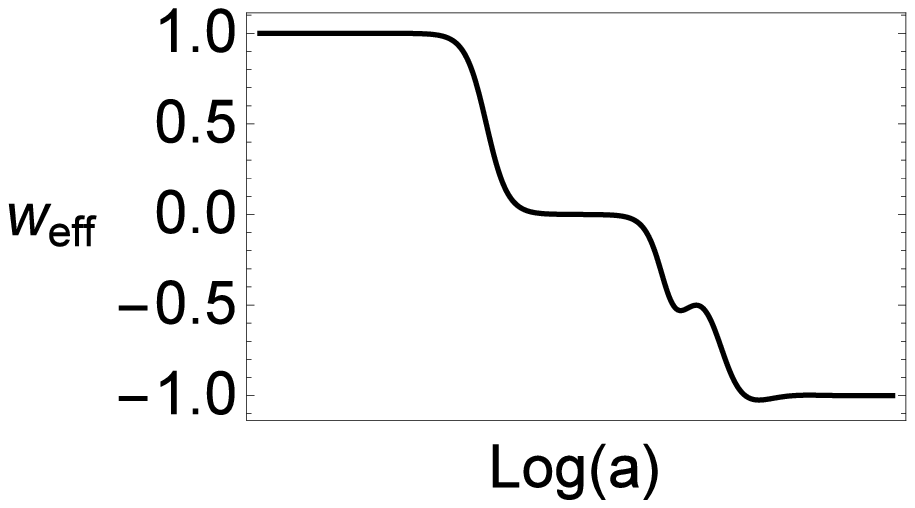}
\hfill
\includegraphics[width=0.48\textwidth]{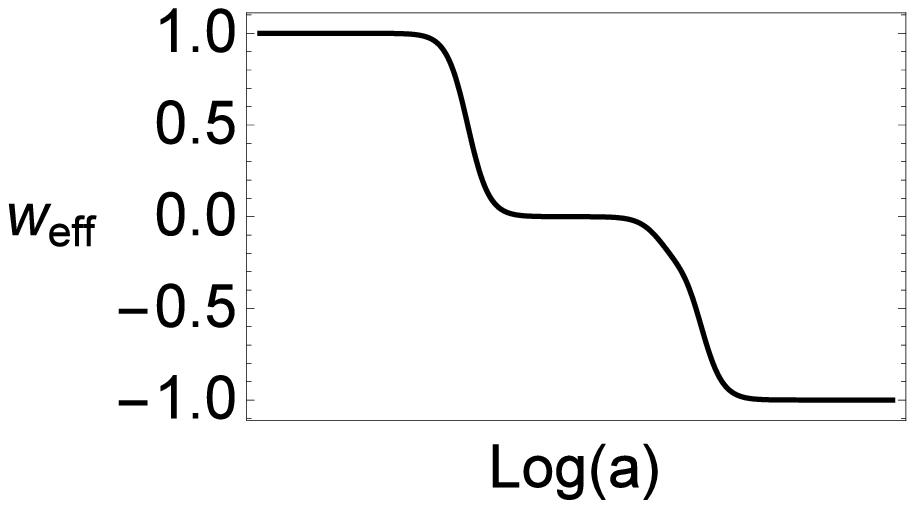}
\caption{$w_{\rm eff}$ of model B of the dynamical system. Left panel: $w=0$, $\lambda=2$, $\xi=1$. Right panel: $w=0$, $\lambda=2$, $\xi=-1$.}
\label{3dxi1weff}
\end{figure}

To better expose the dark matter to dark energy transition at late times, we consider the behaviour of the effective equation of state for particular trajectories within the phase space of Model~B and compare its qualitative properties. For this purpose, we look at the dynamical evolution of $w_{\rm eff}$ for different choices of the parameters of Model~B; see Fig.~\ref{3dxi1weff}. We see that in this model we always find physical trajectories starting with the usual scalar field kinetic energy dominated epoch and then evolving through a matter dominated phase, after which they approach the late time dark energy dominated attractor. Note that in this case there is the possibility of dynamically crossing the phantom barrier, after which the effective equation of state approaches $w_{\rm eff}=-1$ from below; see the left panel of Fig.~\ref{3dxi1weff}. However, this behaviour is parameter dependent. For instance, when choosing $\xi=-1$ the value $w_{\rm eff}=-1$ is approached from above. Note also that a non-monotonic transition from matter to dark energy domination can be obtained, as shown again in the left panel of Fig.~\ref{3dxi1weff}. All this shows the interesting phenomenology present in Model~B and suggests further investigations with other similar models.

We might for example compare the coupling appearing in Eqs.~(\ref{048b}) and (\ref{049b}) with the one usually considered in the dynamics of early-universe bubble nucleation arising from first-order phase transitions; see e.g.~\cite{Ignatius:1993qn}. In those models the coupling vector $Q_\mu$ between the scalar field and the surrounding matter fluid is provided by $Q_\mu \propto U^\nu \partial_\nu\phi \partial_\mu\phi$ and it is motivated by thermodynamical properties of electro-weak physics, such as the temperature dependence of the Higgs potential under renormalization at two or more loops. This coupling vector is different from the one arising in Eqs.~(\ref{048b}) and (\ref{049b}) which assumes the form $Q_\mu \propto \nabla_\nu U^\nu \partial_\mu\phi$ (and also retains a quite general dependence\footnote{In \cite{perturbations} it will be shown that only models where $f \propto 1/n$ are generally free from instabilities at the cosmological perturbations level.} on the coupling function $f$). Unfortunately it seems not easy to obtain the coupling $Q_\mu \propto U^\nu \partial_\nu\phi \partial_\mu\phi$ from the scalar-fluid variational approach considered in this work, even if higher-order operator such as $\partial_\mu\phi\partial^\mu\phi$ are taken into account in the Lagrangian. Nevertheless, at a more phenomenological level, the coupling vector $Q_\mu \propto \nabla_\nu U^\nu \partial_\mu\phi$ can be used to describe new dissipative interactions between the expanding bubble and the surrounding fluid, which might give rise to a different dynamics in the early universe and possibly to new observational signatures of first-order phase transitions. These topics might well represent future applications of the variational formalism developed here.

Finally we briefly comment on the appearance of a ``fifth'' force $f^\mu$ acting on the matter fluid due to the interaction with the scalar field. From Eq.~(\ref{997}) (or equivalently from Eq.~(\ref{998})) it is possible to derive the geodesic equation for the fluid which reads
\begin{align}
  \frac{dU^\mu}{d\tau} + \Gamma^\mu_{\alpha\beta} U^\alpha U^\beta = -\frac{h^{\mu\nu}}{\rho+p+p_{\rm int}} \left[ \partial_\nu p + \partial_\nu p_{\rm int} -  n^2 \frac{\partial f}{\partial n} \nabla_\lambda U^\lambda  \nabla_\nu\phi \right] = f^\mu \,,
  \label{995}
\end{align}
where $p_{\rm int}$ is defined in Eq.~(\ref{996}). Note that this fifth force is orthogonal to the fluid flow, $f^\mu U_\mu =0$, in agreement with the relativistic definition of 4-force. The presence of a non-vanishing force on the right hand side of the geodesic equation, even for a dust fluid ($p=0$), implies that in general the motion of matter particles will be non geodesic and that the equivalence principle will be violated. If the scalar field interacts with baryonic matter, then Solar System experiments set strong constraints on the magnitude of this fifth force \cite{Will:2014xja}, which must then be somehow negligible at small scales. However since it depends on local quantities such as the scalar field, the particle number density (or energy density) and the entropy density, a screening mechanism, similar to the well known chameleon mechanism \cite{Khoury:2003aq}, could be at play. The interesting and original feature of the non geodesic force (\ref{995}) is that it also depends on the fluid 4-velocity $U^\mu$. This suggests the possibility of finding new screening mechanisms for the scalar field which hide its effects wherever the matter velocities are relativistically small, such as in the Solar System. Note that this cannot be achieved neither with the usual chameleon theories nor with the algebraic couplings of Part~I, since in that cases the resulting fifth force does not depend on the fluid 4-velocity. Unfortunately the resulting analysis at small scales is complicated by this dependence on the matter velocities and cannot be easily performed following the original chameleon work, as it has been done for the algebraic couplings in Part~I. The study of these new screening mechanisms depending on the matter fields velocities, as well as their phenomenological effects, falls well beyond the scope of the present paper and will be left for future considerations.

In conclusion the variational approach outlined in this work is very powerful because it always allows us to arrive at covariant theories where the background and the perturbations can be studied consistently. It can also represent the starting point for more radical interactions between a scalar field and a matter fluid. For instance, one could attempt to identity the scalar field $\phi$ with either of the two Lagrange multipliers $\varphi$ or $\theta$. In particular the first one would allow for fluid's particle creation/annihilation, while the second one would correspond to the scalar field affecting the entropy of the system. Clearly, there is a large number of possible realisations whose consequences and applications could be studied in the future using different techniques, such as dynamical systems and perturbations. The investigation of cosmological perturbations constitutes the next logical step for further understanding the physics of these models and to better compare them against observations.

\end{document}